\documentclass[runningheads]{llncs}

\usepackage[utf8]{inputenc}
\usepackage[T1]{fontenc}
\usepackage{graphicx}

\usepackage{booktabs}
\usepackage{tabularx}
\usepackage{acronym}
\usepackage[table]{xcolor}
\usepackage{float}
\usepackage{booktabs}
\usepackage{float}
\usepackage{amssymb}
\usepackage{booktabs}
\usepackage{tabularx}

\begin{document}

\begin{center}
\small\itshape \textbf{This is a preprint version. The final version will appear in the proceedings of PROFES~2026.}
\end{center}

\title{From Data Quality to Data Trustworthiness in AI-Driven Software-Intensive Embedded Systems: A Systematic Review}
\titlerunning{From Data Quality to Data Trustworthiness in AI-Driven SIES}

\author{Hariharan Gopinath\inst{1}\thanks{Corresponding author.} \and
Jan Bosch\inst{1,2} \and
Helena Holmström Olsson\inst{3}}

\authorrunning{H. Gopinath et al.}
\institute{
Chalmers University of Technology, Department of Computer Science and Engineering, Gothenburg, Sweden \\
\email{hargop@chalmers.se, jan.bosch@chalmers.se}
\and
Eindhoven University of Technology, Department of Mathematics and Computer Science, Eindhoven, Netherlands \\
\and
Malm\"o University, Department of Computer Science and Media Technology, Malm\"o, Sweden \\
\email{helena.holmstrom.olsson@mau.se}
}
\maketitle

\begin{abstract}
AI-driven software-intensive embedded systems (SIES) depend on data during development and operation. Poor data quality (DQ) may weaken data trustworthiness, yet evidence about its effects and lifecycle management remains fragmented. This systematic literature review synthesizes 27 primary studies published from January 2020 to February 2026. It examines DQ concerns and evidence linking them to reduced data trustworthiness, along with assessment and management methods. The review identifies 16 concerns organized as Source Quality and Content Quality. Studies provide DQ-based scores and evidence of effects on model or decision performance. Synthesis suggests upstream source or handling problems may become embedded in training baselines and remain difficult to detect at runtime. Eight method categories were identified. Runtime monitoring is the largest category, while upstream and training-time controls remain poorly integrated with runtime assurance. No study covers the full lifecycle. Data trustworthiness supports reliable SIES decisions but is one component of AI-system trustworthiness.

\end{abstract}

\section{Introduction}
\label{sec:introduction}

The rapid integration of artificial intelligence (AI) into software-intensive embedded systems (SIES), including autonomous vehicles, Internet of Things (IoT) applications, and safety-critical cyber-physical systems (CPS), has changed how operational decisions are made \cite{TowardstrustworthyAI,DQinIndustrial4.0}. AI components now generate perception outputs and predictive-maintenance recommendations in settings where failure can have physical consequences. These systems depend on training data and continuous streams of sensor, environmental, and contextual data. When DQ degrades because data are incomplete, noisy, drifting, delayed, or maliciously altered, the basis for relying on AI-driven classifications or predictions also degrades.

Trustworthy AI is often framed through model-centric properties such as explainability, interpretability, and transparency \cite{TowardstrustworthyAI}. In SIES and CPS, it also depends on whether training and runtime data reflect operational conditions and have a defensible history \cite{DQinIndustrial4.0}. Dynamic streams can be affected by noise, distribution shift, sensor degradation, and cyberattacks \cite{dynamicDQ}. Traditional DQ frameworks do not explicitly address how data-quality
failures may propagate through learning-enabled operational systems \cite{batini_methodologies_2009}.

Moreover, AI-specific data problems, such as distributional shift, noisy labels, class imbalance, corrupted annotations, and training-data poisoning \cite{whang2022datacollectionqualitychallenges}, arise during both development and operation. They are further complicated by streaming data, monitoring feedback, retraining, and redeployment \cite{dynamicDQ}. The literature uses overlapping terminology \cite{zhou_survey_2024}, mixes conventional DQ dimensions with AI-specific failure modes, and seldom states what evidence justifies reliance on data for a particular task and context. To address this gap, we systematically review 27 primary studies published between January 2020 and February 2026. The review contributes a lifecycle mapping of 16 data-quality concerns, a synthesis of the evidence linking these concerns to data trustworthiness, and a classification of eight assessment and management approaches. We treat data trustworthiness as one component of overall AI-system trustworthiness rather than as evidence of system trustworthiness by itself. The review is guided by the following research questions:

\noindent\textbf{RQ1:} Which data-quality concerns are reported in AI-driven SIES, and where do they arise across the data lifecycle?

\noindent\textbf{RQ2:} What evidence links the data-quality concerns to reduced data trustworthiness for a specified AI task or decision?

\noindent\textbf{RQ3:} How are the data-quality concerns assessed and managed, and which lifecycle gaps remain?

The rest of this SLR is organized as follows. Section~\ref{sec:background} presents the background and related work. Section~\ref{sec:protocol} describes the review methodology. Section~\ref{sec:results} presents the findings and answers to the research questions. Section~\ref{sec:discussion} discusses the insights from the results. Section~\ref{sec:threats} outlines the threats to validity. Finally, Section~\ref{sec:conclusion} presents the conclusion and future work.

\section{Background and Related Work}
\label{sec:background}

\subsection{Data Quality and Data Trustworthiness}
\label{sec:dq_trustworthiness}

Data quality has traditionally been defined in relation to how well data can be used and how accurately it represents reality. Wang and Strong emphasize fitness for use, whereas Wand and Wang relate DQ to correspondence between real-world states and their representation in information systems~\cite{wang_beyond_1996,wandwang}. A DQ dimension is a measurable aspect of this fitness, such as accuracy, completeness, timeliness, or consistency~\cite{wang_beyond_1996,batini_methodologies_2009}. Established frameworks organize these dimensions differently: Wang and Strong distinguish intrinsic, contextual, representational, and accessibility categories, while ISO/IEC 25012 uses inherent and system-dependent characteristics~\cite{wang_beyond_1996,iso25012}. In this review, \textit{data-quality concern} is an umbrella term covering a DQ dimension, data-assurance property, data-related failure mode, or data-model condition that may reduce the suitability of data for an intended AI task.

AI-driven software-intensive embedded systems (SIES) introduce additional data-related concerns across development and operation. During model development, training data may contain class imbalance, noisy or corrupted labels, insufficient coverage, or deliberately poisoned samples~\cite{whang2022datacollectionqualitychallenges}. During operation, continuously generated data may be affected by sensor degradation, environmental noise, out-of-distribution inputs, or changes in the underlying data distribution~\cite{dynamicDQ,zhou_survey_2024}. Such conditions can reduce model or decision reliability and weaken the evidence available for relying on AI outputs. Data quality, data trustworthiness, human trust, and AI-system trustworthiness are related but distinct. Data quality concerns measurable properties of data. Human trust is a stakeholder's willingness to rely on a system under uncertainty~\cite{trustandtrustworthiessinAI}. AI-system trustworthiness has a broader scope and also depends on model behaviour, software and hardware, governance, human oversight, and the operating environment~\cite{trustandtrustworthiessinAI,TowardstrustworthyAI}.

Prior research defines data quality through fitness for use and correspondence between recorded data and real-world conditions~\cite{wang_beyond_1996,wandwang} and treats data and AI trustworthiness as contextual, evidence-based, and subject to reassessment across the lifecycle~\cite{10,28,29}. Based on these foundations, we formulate the following literature-informed working definition for this review: \textit{Data trustworthiness is the context- and time-dependent extent to which verifiable evidence justifies relying on data for a specified AI task or decision.} Adequate data quality alone does not establish human trust, system safety, or overall AI-system trustworthiness~\cite{14,28}.

\subsection{Related Work and Research Gap}
\label{sec:related_work}

Existing studies and reviews address different parts of the relationship between data quality and trustworthiness. Goknil et al.\ review DQ challenges and approaches in industrial cyber-physical and IoT systems, providing a domain-specific account of established DQ dimensions and practices~\cite{DQinIndustrial4.0}. Zhou et al.\ examine DQ dimensions and assessment tools for machine learning more broadly but do not focus on the lifecycle and operational constraints of SIES~\cite{zhou_survey_2024}. Schwabe et al.\ connect data quality to trustworthy AI in medicine, where the application context, data types, and assurance requirements differ from those of embedded and cyber-physical systems~\cite{metric-framework_2024}.

Related trust research also adopts different units of analysis. Bach et al.\ examine human trust in AI-enabled systems from a human-computer interaction perspective~\cite{bach_systematic_2024}, while lifecycle-oriented trustworthy-AI research addresses system-level properties and governance~\cite{TowardstrustworthyAI}. Ouattara et al.\ consider the trustworthiness of AI training datasets but focus primarily on training data rather than the complete development and operational lifecycle~\cite{trust_training_dataset}.

The related work addresses individual parts of this problem but does not jointly synthesize (i) the DQ concerns reported across the lifecycle of AI-driven SIES, (ii) the evidence linking these concerns to reduced data trustworthiness, and (iii) the methods used to assess and manage them. This review addresses the gap by examining data trustworthiness across requirements, collection and ingestion, preparation and curation, training and evaluation, and deployment and monitoring, including feedback loops.

\section{Review Methodology}
\label{sec:protocol}

\begin{table}[H]
\centering
\caption{Normalized concept blocks used to construct database-specific queries}
\label{tab:slr_search_string}
\scriptsize
\setlength{\tabcolsep}{4pt}
\renewcommand{\arraystretch}{1.3}
\begin{tabularx}{\textwidth}{@{} p{2cm} X @{}}
\toprule
\textbf{Concept} & \textbf{Search Terms} \\
\midrule
Data evidence and problems
& \texttt{("data quality" OR "quality of data" OR "data monitoring" OR "data assessment")} \\

AI / ML 
& \texttt{("machine learning" OR "deep learning" OR "artificial intelligence" OR "AI" OR "ML" OR "reinforcement learning" OR "autonomous system*")} \\

Trust / Trustworthiness
& \texttt{("data trust*" OR "trust in data" OR "confidence in data" OR "trustworth*" OR "model trust")} \\

Application
& \texttt{("software-intensive embedded system*" OR "embedded system*" OR "cyber-physical system*" OR "industrial AI" OR "autonomous system*" OR "safety-critical system*" OR "real-time system*" OR "industrial IoT" OR "edge AI")} \\
\bottomrule
\end{tabularx}
\end{table}

This review followed the guidelines of Kitchenham and Charters~\cite{slr_guidelines} and used the CIMO framework~\cite{CIMO} to define the scope and search strategy. The search combined four concept groups: data quality, AI/ML, trust or trustworthiness, and software-intensive embedded or cyber-physical systems, as shown in Table~\ref{tab:slr_search_string}. IEEE Xplore, ACM Digital Library, and Scopus were searched for studies published between January 2020 and February 2026. The lower bound reflects the growth of research on AI-related challenges in industrial CPS from 2020 onward~\cite{windmann_artificial_2024}. Forward and backward snowballing were then used to identify additional studies. All records found through snowballing were assessed using the same eligibility and quality-appraisal procedures as the database results. Zotero was used to manage references, remove duplicates, and organize the
retrieved studies.

\textbf{Eligibility and quality appraisal:}

Studies were eligible if they were peer-reviewed, written in English, published within the review period, and examined data-quality concerns or data-related failures relevant to at least one RQ in an AI-driven embedded, CPS, edge, or IoT context. SIES relevance required an AI-enabled system involving a physical process or sensor stream, embedded or edge execution, real-time or resource constraints, or an explicit industrial IoT, autonomous-system, or digital-twin setting. We excluded secondary, inaccessible, duplicate or superseded studies, as well as cloud-only, traditional database, and model- or security-only studies without relevant AI data-quality evidence.

Eligibility and study quality were assessed during complete full-text review using ten criteria. Q1-Q8 assessed methodological and evidential quality: (Q1) clarity of the research objective; (Q2) description of the study context; (Q3) definition and operationalization of the examined data-quality properties or methods; (Q4) appropriateness and clarity of the research methodology; (Q5) discussion of limitations or threats to validity; (Q6) evidential support for the claimed relationship between the examined data condition and trustworthiness, reliability, decision confidence, or another reliance-related outcome; (Q7) support for the reported findings; and (Q8) consistency between the results and conclusions. Q9 and Q10 assessed eligibility: (Q9) relevance to an AI-driven embedded, CPS, edge, or IoT context; and (Q10) contribution to at least one research question. 
Each criterion was scored as 1, 0.5, or 0, giving a maximum appraisal score of 10, and retained studies also required a total score of at least 6. Studies scoring below 6, corresponding to less than 60\% of the maximum score, were excluded. Retained studies scoring 6-7.5 were classified as moderate-quality studies, while those scoring 8-10 were classified as high-quality studies. The assessment procedure and scoring scheme were informed by Kitchenham and Charters~\cite{slr_guidelines}.

\textbf{Data extraction and synthesis:}
A standardized form recorded bibliographic and study details, application domain, AI/ML technique, evaluation context, DQ terminology, reported links to trustworthiness or reliability, lifecycle location, and assessment or management methods. The synthesis followed six steps. We (1) extracted relevant terms, definitions, measures, and supporting passages using each study's terminology, (2) compared and analyzed them with established DQ vocabularies~\cite{wang_beyond_1996,batini_methodologies_2009}, (3) merged terms only when their definitions and operational roles matched, (4) classified each concern by type, (5) grouped concerns by their main evidential focus, and (6) grouped assessment and management methods by purpose and lifecycle location. Cross-cutting concerns were assigned to their point of origin, while downstream effects were retained in the qualitative synthesis. The resulting analytical views and method categories, presented in Sections~\ref{subsec:rq1} and~\ref{subsec:rq3}, emerged from the cross-study synthesis and were not predefined selection criteria.

For lifecycle mapping, we consolidated the nine-stage machine-learning workflow of Amershi et al.~\cite{data_AI_lifecycle} into five analytical locations: \textit{1) Requirements, 2) Collection and ingestion, 3) Preparation and curation, 4) Training and evaluation, and 5) Deployment and monitoring. }Preparation and curation cover data cleaning, labelling, and feature engineering. Deployment and monitoring include runtime operation. Activities spanning several stages are marked as \textit{across lifecycle}. Retraining, correction, and redeployment are treated as feedback loops rather than separate linear stages.

\section{Results}
\label{sec:results}
\subsection{Study Selection Outcomes}
\label{sec:SLRoutcome}

\begin{figure}[H]
    \centering
    \includegraphics[width=1\textwidth]{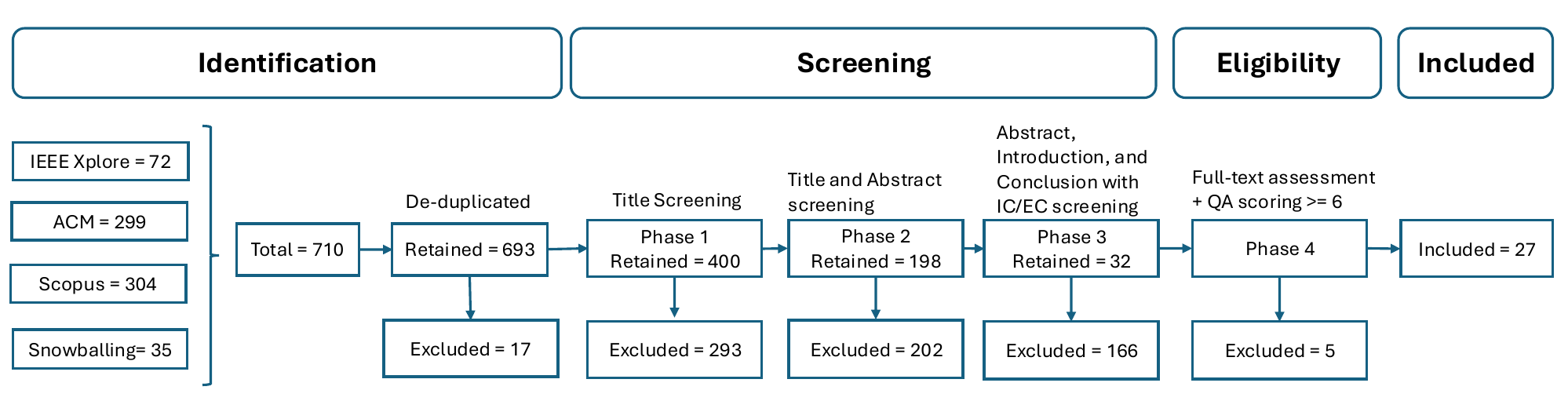}
    \caption{PRISMA flow diagram for study selection}
    \label{fig:prisma}
\end{figure}

 Figure~\ref{fig:prisma} summarizes the study-selection process. After
deduplication, 693 records remained. Title screening retained 400 records, and
title and abstract screening retained 198. Focused screening using the inclusion
and exclusion criteria retained 32 studies for complete full-text assessment.
These studies were read in full and evaluated using the ten assessment criteria,
with Q1-Q8 addressing methodological and evidential quality and Q9-Q10
addressing relevance to the SIES scope and contribution to the research
questions. Five studies scored below the minimum threshold of 6 and were
excluded. The final synthesis, therefore, included 27 studies: 17 scored
6-7.5 and were classified as moderate-quality studies, while 10 scored 8-10 and were
classified as high-quality studies. A sensitivity check including the five studies below the cutoff did not change
the 16 concerns, the SQ/CQ structure, the eight method categories, or the main
conclusions.

\begin{table}[H]
\centering
\caption{Primary studies included in the synthesis}
\label{tab:primary-studies}
\scriptsize
\setlength{\tabcolsep}{2pt}
\renewcommand{\arraystretch}{1.0}
\begin{tabularx}{\textwidth}{@{} *{9}{>{\centering\arraybackslash}X} @{} }
01 \cite{1} & 02 \cite{2} & 03 \cite{3} & 04 \cite{4} & 05 \cite{5} & 06 \cite{6} & 07 \cite{7} & 08 \cite{8} & 09 \cite{9} \\
10 \cite{10} & 11 \cite{11} & 12 \cite{12} & 13 \cite{13} & 14 \cite{14} & 15 \cite{15} & 16 \cite{16} & 17 \cite{17} & 18 \cite{18} \\
19 \cite{19} & 20 \cite{20} & 21 \cite{21} & 22 \cite{22} & 23 \cite{23} & 24 \cite{28} & 25 \cite{29} & 26 \cite{30} & 27 \cite{32} \\
\end{tabularx}
\end{table}

\subsection{RQ1 - Data-Quality Concerns Across the AI Lifecycle}
\label{subsec:rq1}

The cross-study synthesis identified 16 recurring data-related concerns, including conventional DQ dimensions, data-assurance properties, failure modes, and data-model conditions. Table~\ref{tab:dq_lifecycle} maps these concerns across the AI lifecycle. To organize the evidence, we derived two complementary analytical views: \textit{Source Quality (SQ)} and \textit{Content Quality (CQ)}. SQ concerns the origin,
acquisition, provenance, sensor condition, and handling of data. CQ concerns whether data values, labels, coverage, and distributions remain suitable for the intended task, context, and time. SQ and CQ are complementary analytical views rather than mutually exclusive or hierarchical categories. A study may contribute evidence to both views; therefore, the SQ and CQ study counts are not additive. Terminology was normalized by definition and operational role. For example, \textit{"erroneous values"} \cite{4}, \textit{"corrupted data"} and \textit{"noise"} \cite{2} were grouped under \textit{"accuracy"} when they referred to incorrect or distorted measurements. A source-related problem can later appear as a content-quality problem. For example, sensor degradation may result in inaccurate values or distribution drift. These views do not replace established DQ frameworks~\cite{wang_beyond_1996,batini_methodologies_2009}. Instead, they are an author-derived synthesis for organizing the forms of evidence reported in the reviewed studies.
\begin{table}[H]
\centering
\caption{Data-related concerns by analytical view and lifecycle stage. Slash-separated terms represent normalized concern families and are not
necessarily exact synonyms.}
\label{tab:dq_lifecycle}
\scriptsize
\setlength{\tabcolsep}{4pt}
\renewcommand{\arraystretch}{1.1}
\begin{tabularx}{\textwidth}{@{} c p{3.6cm} p{3.8cm} p{3.2cm} @{}}
\toprule
\textbf{Code} & \textbf{Concern} & \textbf{Lifecycle location} & \textbf{Studies} \\
\midrule
\multicolumn{4}{c}{\cellcolor{gray!20}\textbf{Source quality (SQ): origin, acquisition, and handling evidence}} \\
\midrule
SQ1 & Source identity / authenticity / provenance & Collection and ingestion & \cite{10,15,17,19,22,29,30} \\
SQ2 & Traceability / lineage / versioning & Across lifecycle & \cite{3,4,6,7,10,15,17,22,32} \\
SQ3 & Data or model-update tampering / poisoning & Collection and ingestion; Training and evaluation; Feedback loop & \cite{7,9,10,14,15,17} \\
SQ4 & Data integration and curation quality & Collection and ingestion; Preparation and curation & \cite{4,6,10,14,20,22,32} \\
SQ5 & Sensor acquisition condition / degradation & Collection and ingestion; Deployment and monitoring & \cite{7,16,17,22,30} \\
\midrule
\multicolumn{4}{c}{\cellcolor{gray!20}\textbf{Content quality (CQ): fitness for task, context, and time}} \\
\midrule
CQ1 & Data label / annotation quality & Preparation and curation & \cite{3,12,17} \\
CQ2 & Completeness / missingness & Collection and ingestion; Preparation and curation & \cite{2,4,6,7,14}\newline\cite{20,22,29,30,32} \\
CQ3 & Accuracy / noise / erroneous values & Collection and ingestion; Preparation and curation; Deployment and monitoring & \cite{2,4,6,7,10,12,14}\newline\cite{16,17,20,22,23,30,32} \\
CQ4 & Timeliness / currency / age of information & Collection and ingestion; Deployment and monitoring & \cite{4,14,22,28,30} \\
CQ5 & Consistency / representation & Preparation and curation & \cite{4,6,14,20,22,32} \\
CQ6 & Anomalies / outliers & Preparation and curation; Deployment and monitoring & \cite{4,6,10,12,14}\newline\cite{17,20,22,30,32} \\
CQ7 & Validity / conformity & Preparation and curation & \cite{4,6,14,20,32} \\
CQ8 & Uniqueness / duplication & Preparation and curation & \cite{4,6,14,22} \\
CQ9 & Data representativeness / operational coverage & Collection and ingestion; Training and evaluation & \cite{3,7,12,16,23} \\
CQ10 & Out-of-distribution (OOD) data & Deployment and monitoring & \cite{1,2,5,8,11}\newline\cite{12,13,16,18,23} \\
CQ11 & Distribution shift / concept drift & Deployment and monitoring & \cite{2,5,12,16,17}\newline\cite{18,21,23,32} \\
\bottomrule
\end{tabularx}
\end{table}
\textbf{Source-quality view:} SQ items appear in 16 of the 27 studies and concentrate at collection and ingestion, preparation and curation, and training and evaluation. Provenance and traceability concern whether origin and transformations can be examined.  Data integration and curation quality are illustrated by Fischer et al.~\cite{14}, where records failing integration checks receive a lower trust score before inference. Deliberate training-data poisoning is treated as a handling failure, while sensor degradation is treated as an acquisition condition. Their downstream symptoms, such as inaccurate values or drift, are represented in the CQ view. A degrading sensor is therefore an SQ acquisition problem, while its distorted readings can appear downstream as CQ accuracy loss or drift.

\textbf{Content-quality view:} CQ items appear in 24 studies. They include established dimensions such as accuracy, completeness, timeliness, and consistency. They also include training-time concerns and representativeness and runtime conditions such as OOD inputs and distribution shift. OOD denotes an input outside the learned distribution, while distribution shift denotes a systematic change in the data-generating process over time \cite{zhou_survey_2024}. Normalization followed the operational measure, not the authors' label alone. For example, Bayram et al.~label a distribution test as ``timeliness'' and Jensen--Shannon divergence as ``skewness.'' We treat these as distributional conditions rather than conventional freshness or skewness \cite{32}. The same data can be adequate for one task or time and inadequate for another. RQ2 therefore examines what evidence connects the reported SQ/CQ concerns to a reduced basis for reliance, while RQ3 examines the methods and remaining lifecycle gaps.

\subsection{RQ2 - Evidence Linking Data-Quality Concerns to Data Trustworthiness}
\label{subsec:rq2}
The reviewed studies provide three forms of evidence linking the DQ concerns to reduced data trustworthiness. The findings are interpreted using the working definition of data trustworthiness introduced in Section~\ref{sec:background}.

\textbf{Score-based evidence.}
Several studies convert DQ measurements into scores used to accept, reject, or gate data. Fischer et al.~\cite{14} lower the score of records that fail cross-system checks, although their formula is not validated end to end. A power-system study proposes a DQ impact coefficient but does not empirically calibrate its thresholds~\cite{20}. Byabazaire et al.~first propose a three-stage trust framework~\cite{29} and later evaluate it experimentally~\cite{30}. Under injected noise, the DQ-derived score decreases as prediction error increases, with a reported Pearson correlation of $-0.95$~\cite{30}. Because the score is constructed from DQ measures, this result shows that it tracks model error in that experiment. It does not establish an independent causal effect on human trust or overall system trustworthiness.

\textbf{Performance and safety evidence.}
Other studies examine the effects of DQ problems on technical outcomes. In Kabashkin's simulated hydraulic-monitoring case, progressive sensor drift reduces recall from 91.1\% to 78.3\% across federated-learning rounds, while federated unlearning restores much of the lost performance~\cite{17}. Studies of OOD inputs also show that neural classifiers may remain confident when processing unfamiliar data~\cite{1,5,8,11,13,18}. Without rejection or fallback mechanisms, this confidence provides a weak basis for relying on the output. However, an OOD input does not necessarily cause a safety violation. Lin et al.~therefore monitor the safety property directly under distribution shift rather than treating OOD detection as sufficient~\cite{2}. Overall, these studies measure error, recall, confidence, rejection, fallback, or safety alarms, but do not directly evaluate operator trust.

\textbf{Cross-study propagation pattern.}
Across the studies, we derived a recurring cross-study risk: a source or handling problem introduced before training may become part of the training data, reference distribution, or model. Sample et al.~describe this risk for poisoned CPS training baselines, while later studies show that corrupted contributions and sensor drift may persist until explicit monitoring or correction is applied~\cite{7,17}. A runtime monitor using the same affected baseline may then miss the original problem or detect only a downstream symptom. This pattern is not universal. Independent provenance evidence, trusted reference data, sensor redundancy, and training-time checks can interrupt the pathway. Provenance establishes data origin and history, but does not by itself prove that the recorded values are correct~\cite{15,19}.

Taken together, the evidence supports task- and context-specific decisions to accept, reject, monitor, or restrict data. However, it does not show that DQ alone determines human trust, system safety, or the trustworthiness of the complete AI system.

\subsection{RQ3 - Assessment, Management, and Lifecycle Gaps}
\label{subsec:rq3}

RQ3 examines how DQ concerns are assessed and managed and where lifecycle gaps remain. Table~\ref{tab:rq3-unified}  groups eight method categories by role and lifecycle location. Category roles and lifecycle locations describe their combined coverage. A study may appear in several categories but need not perform every listed role.

\textbf{How DQ concerns are assessed. } Assessment combines requirements and evidence checks, profiling, provenance verification, scoring, runtime monitoring, drift detection, and freshness measures. Runtime monitoring and response is the largest category~\cite{1,2,5,8,10,11,12,13,16,18,23}. It uses certainty scores~\cite{1}, reconstruction error~\cite{5}, activation coverage~\cite{8}, calibrated uncertainty~\cite{13,18}, and distance measures~\cite{23} to judge support for inputs or predictions. Composite scores assess different units. Fischer et al. score supply-chain records~\cite{14}, An proposes an enterprise DQ impact coefficient~\cite{20}, Byabazaire et al. first propose a three-stage framework that combines DQ measures into a trust score for shared IoT data~\cite{29}. They later evaluate it under injected noise and relate the score to prediction error~\cite{30}. Bayram et al. score training windows and assess how acceptance thresholds affect prediction accuracy~\cite{32}.

\textbf{How detected problems are managed. }The response depends on the method and lifecycle stage. Profiling methods repair, remove, or impute missing, noisy, anomalous, or invalid data~\cite{4,6,10,22}. Runtime methods may raise an alarm, reject a prediction, restrict a data source, or activate a fallback. Drift-detection methods can trigger model updating, retraining, or redeployment~\cite{12,17,21,32}. Training and feedback methods attempt to prevent poisoning or remove harmful federated contributions after they are discovered~\cite{7,14,17}. Provenance and auditability methods preserve information about data origin and processing history, although they cannot by themselves establish that the recorded values are correct~\cite{15,19}.

\textbf{Which lifecycle gaps remain. }Runtime monitoring is the largest category, but upstream and training-time controls remain weakly connected to it. No included study covers the complete lifecycle. Requirements and evidence assurance~\cite{3} and freshness-aware delivery~\cite{28} each appear once. Source identity and auditability are more common but remain weakly linked to training and runtime assurance~\cite{4,6,7,10,15,17,19}. Only three address training-data or model-update integrity~\cite{7,14,17}. One study reports communicating an OOD condition to an operator, but operator interpretation and response are not evaluated~\cite{16}. Resource findings are difficult to compare because costs range from sub-millisecond checks to substantial inference-time increases across platforms and workloads~\cite{6,8,11}. Memory, energy, and communication costs are rarely measured while end-to-end latency is reported in one industrial study~\cite{32}.

The methods cover several lifecycle locations, but their connections remain limited. No included study integrates upstream evidence, training-time controls, runtime responses, and feedback into a lifecycle-wide assurance approach.

\begin{table}[t]
\caption{DQ assessment and management methods}
\label{tab:rq3-unified}
\centering
\scriptsize
\setlength{\tabcolsep}{3pt}
\renewcommand{\arraystretch}{1.04}

\begin{tabularx}{\textwidth}{@{} X p{2.1cm} p{2.4cm} @{}}
\toprule
\textbf{Method, role, and purpose} &
\textbf{Lifecycle location} &
\textbf{Studies} \\
\midrule

\textbf{Requirements and evidence assurance (A)}$^{*}$:
defines DQ requirements and assesses evidence against them &
Requirements; Preparation and curation &
\cite{3} \\

\textbf{Profiling and curation (A/M)}:
detects, cleans, repairs, removes, or imputes deficient data &
Preparation and curation; Deployment and monitoring &
\cite{4,6,10,22} \\

\textbf{Composite scoring and gating (A/M)}:
scores data, sources, or training windows and may control their use &
Across lifecycle &
\cite{14,20,29,30,32} \\

\textbf{Source identity, provenance, and auditability (A/M)}:
records or verifies data origin, lineage, versions, transformations,
and update history &
Collection and ingestion; Across lifecycle &
\cite{4,6,7,10,15,17,19} \\

\textbf{Runtime monitoring and response (A/M)}:
detects unsupported inputs or predictions and may trigger alarms,
rejection, or fallback &
Deployment and monitoring &
\cite{1,2,5,8,10,11}
\cite{12,13,16,18,23} \\
\textbf{Drift detection and adaptation (A/M)}:
detects distribution or concept drift and may trigger updating,
retraining, or redeployment &
Deployment and monitoring; Feedback loop &
\cite{12,17,21,23,32} \\

\textbf{Training-data and model-update integrity (A/M)}:
detects, prevents, filters, or removes corrupted training data
or model updates &
Preparation and curation; Training and evaluation; Feedback loop &
\cite{7,14,17} \\

\textbf{Freshness-aware data delivery (A/M)}$^{*}$:
measures data freshness and selects data under resource constraints &
Collection and ingestion; Deployment and monitoring &
\cite{28} \\

\bottomrule
\end{tabularx}

\vspace{1mm}
\raggedright
\scriptsize
A: assessment; M: management. Roles and lifecycle locations describe the
combined coverage of the studies in each category. Individual studies may
perform only one role or address only part of the stated lifecycle coverage.
$^{*}$Category represented by one study.
\end{table}

\section{Discussion: From DQ Checks to Lifecycle Assurance}
\label{sec:discussion}

\begin{figure}[H]
    \centering
    \includegraphics[width=1.01\textwidth]{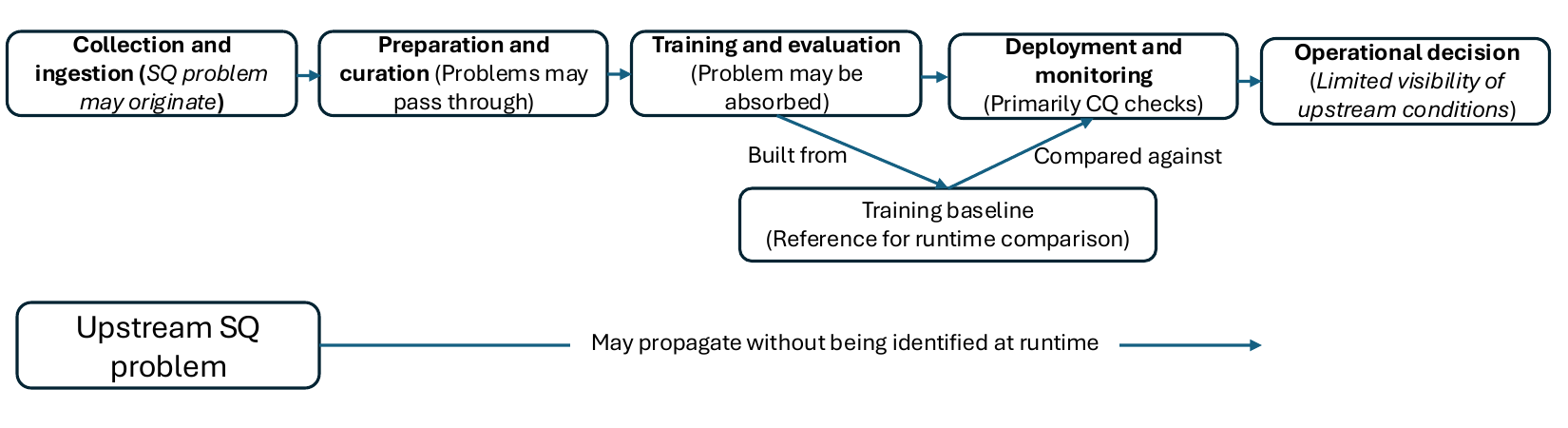}
    \caption{Synthesis of upstream SQ problem propagation. A source or handling problem may pass through data preparation and become part of the training baseline. Runtime monitoring that relies mainly on CQ checks may then miss the original problem or detect only a downstream symptom.}
    \label{fig:propagation}
\end{figure}

The reviewed methods cover several lifecycle stages, although runtime monitoring is the largest method family. Upstream controls for provenance, training integrity, and data handling are less integrated with runtime assurance. Evidence from poisoning and sensor-drift studies suggests that some SQ problems may become embedded before deployment~\cite{7,17}. Figure~\ref{fig:propagation} illustrates this cross-study proposition. It combines evidence from upstream-failure and runtime-monitoring studies and should not be interpreted as an end-to-end causal model validated by one study.

\textbf{Runtime monitoring is not lifecycle assurance.}
An upstream source or handling problem may become part of the training data, reference distribution, or deployed model. Kabashkin’s simulated sensor-drift case illustrates the upstream part of this pathway, where progressive degradation affected federated training before later correction through unlearning~\cite{17}. A runtime monitor that compares new data with an affected baseline may miss the original problem or detect only a later symptom. Independent reference data, provenance checks, sensor redundancy, and training-time controls may help interrupt this pathway. Runtime monitoring therefore remains necessary, but it cannot provide complete evidence about the integrity of earlier lifecycle stages.

This finding extends the existing DQ literature beyond identifying dimensions or monitoring techniques. The main issue is not only whether a runtime input is accurate or in-distribution. It is also whether the reference used to judge that input was produced from sufficiently reliable data and handling processes. The SQ/CQ distinction makes this dependency visible. SQ problems may originate during collection, handling, or training, while their effects may later appear as CQ problems such as inaccurate values, drift, or unsupported predictions.

\textbf{DQ evidence rarely reaches operators.}
Most reported methods act inside the technical pipeline. They gate a data source, reject an input, activate a fallback, or trigger retraining. Only one study reports communicating an OOD condition to an operator in natural language~\cite{16}, and no study evaluates how operators interpret or act on DQ evidence. As a result, the system may detect a problem without giving the operator enough information to understand its cause, severity, or operational consequence. For high-risk systems such as autonomous vehicles, relevant DQ conditions and system responses should therefore be reviewable by the people responsible for operation and oversight. Articles 10, 13, and 14 of the EU AI Act emphasize data governance, transparency, and human oversight~\cite{euaiact2024}. The reviewed studies do not establish a validated operator interface, but they show a clear engineering need to communicate when data are rejected, when fallback is activated, and what evidence supports that action.

\textbf{Data trustworthiness contributes to, but does not determine, system trustworthiness.}
Reliable data cannot compensate for an unsafe model, faulty software, weak governance, or inadequate human oversight. At the same time, a DQ problem does not necessarily result in harm when the system detects the condition and moves to a safe state. Data trustworthiness should therefore be treated as one part of a broader assurance case rather than as evidence of complete system trustworthiness.

The practical question is not whether data are trustworthy in the abstract. It is whether the available evidence supports a specific action under stated conditions. Depending on the evidence, the appropriate action may be to accept the data, monitor them, restrict their use, reject them, or activate a fallback.

\textbf{Engineering implications.}
The synthesis suggests three candidate control points across the lifecycle. First, during data collection and training, assurance should examine data origin, sensor condition, lineage, labels, coverage, and poisoning risks. Second, before deployment and during operation, it should include independent test data, acceptance thresholds, runtime monitoring, and fallback actions. Third, incidents, corrections, and retraining should update the evidence supporting the intended use of the data. These implications are derived from the review and require validation in deployed SIES.

\section{Threats to Validity}
\label{sec:threats}

\textit{Review-process validity.}
The first author conducted study selection, extraction, appraisal, and coding,
while the co-authors reviewed the protocol and final synthesis. Because screening
and coding were not independently replicated, borderline eligibility decisions,
terminology normalization, and SQ/CQ classification may reflect reviewer
judgement. Study-level decisions, scores, and extraction records were retained
for traceability. The 60\% cutoff and score bands were defined for this review;
although the sensitivity check did not change the synthesis, alternative scoring
rules could affect borderline studies.

\textit{Construct and synthesis validity.}
The primary studies use data quality, trust, trustworthiness, reliability, and
confidence inconsistently. Most report technical outcomes, such as error, recall,
uncertainty, or computed scores, rather than human trust or complete-system
trustworthiness. The SQ/CQ classification and Figure~\ref{fig:propagation} are
author-derived syntheses and should not be interpreted as universal or causal
claims.

\textit{Coverage and generalizability.}
The search covered IEEE Xplore, ACM Digital Library, and Scopus and included
English-language peer-reviewed studies published from January 2020 to February
2026. Relevant work may have been missed because of database coverage,
terminology, publication bias, or trust-related search terms. The corpus is
dominated by CPS, industrial IoT, and autonomous systems, limiting
generalizability. Heterogeneous datasets, methods, and outcomes precluded
meta-analysis. Evidence remains limited for foundation models, agents, synthetic
or retrieved data, human feedback, and assurance costs; resource overhead is
reported more often than its effect on training-data quality. The conclusions are
therefore bounded to the learning-enabled SIES represented in the corpus.

\section{Conclusion and Future Work}
\label{sec:conclusion}

This systematic literature review examined data-quality concerns, their relationship to data trustworthiness, and the methods used to assess and manage them in AI-driven SIES. The review identified 16 data-related concerns, organized through complementary source quality and content quality views, and eight categories of assessment and management methods. Runtime monitoring is the most frequently studied area, while the integration of source and training controls with runtime assurance receives less attention.

The synthesis also indicates that an upstream source or handling problem may become part of the training baseline. Runtime monitoring that relies on this baseline may therefore miss the original problem or detect only a downstream symptom. Data trustworthiness supports decisions to accept, monitor, restrict, or reject data, but does not establish overall AI-system trustworthiness. Future work should validate this lifecycle relationship in industrial SIES, evaluate the effectiveness and resource costs of upstream controls, and examine how DQ evidence is communicated to operators and updated through retraining.

\textbf{Declaration of AI-assisted technologies in the writing process:} During manuscript preparation, the first author used QuillBot for language refinement. All suggestions were reviewed and revised by the first author, who takes full responsibility for the final content.

\begin{credits}
\subsubsection{\ackname}
We would like to thank the Software Center companies for partly funding this research.
\end{credits}

\bibliographystyle{splncs04}
\bibliography{references}

\end{document}